# Phenomenological theory of spin-orbit phase transitions

Khisa Sh. Borlakov* and Albert Kh. Borlakov

*North Caucasian State Humanitarian and Technological Academy,*

*36 Stavropolskaya str., Cherkessk, Russia, 369001*

This work is devoted to the logical proof of the Goodenough and Khomskii idea of the existence of spin-orbit transitions in transition magnetic crystals. In agreement with the basics of the Landau theory of phase transitions the phenomenological theory of spin-orbit transitions is constructed. The general scheme of the theory is illustrated by the application to the description of magnetic and structural transformations in the copper ferrite $CuFe_2O_4$.

## 1. Introduction

This paper is the development of the contents of the papers [1-3]. In those works we have shown existence of the cooperative effect of relativistic interactions in magnetic compounds of 3d elements. That idea has been expressed even earlier by a number of authors [4-6] who suggested existence of spin-orbit phase transitions. But in those earlier works (done in late 80s) the idea was not properly formalized and there was neither a phenomenological nor the statistical theory of spin-orbit phase transitions. Now the situation is quite different and the existence of spin-orbit phase transitions has gained wide support. Leading experts in the field (the authors of [7-9] and a number of their followers) offer a variety of microscopic and semi-microscopic models of phase transitions and their applications to specific crystals. However, our phenomenological approach has several big advantages. It allows accounting for the symmetry of all the occurring phases, and its methods are more closely related to empirical studies. In this work, we suggest a phenomenological theory of spin-orbit (or, more generally, relativistic) phase transitions in the spirit of the thermodynamic Landau theory [10]. As an illustration, we show that this theory agrees well with physical effects observed in the copper ferrite $CuFe_2O_4$ in a wide temperature range from above the Curie point $T_c = 730$ K to room temperatures and even lower.

## 2. Symmetry of the magnetic Hamiltonian

Following Van Vleck [11], we write the magnetic Hamiltonian in the form

$$\hat{H} = \hat{H}_{ex} + \hat{H}_{dd} + \hat{H}_{anis} \tag{1}$$

where

$$\hat{H}_{ex} = \sum_{i<k} A_{ik} \vec{S}_i \vec{S}_k \qquad (2)$$

is the exchange Hamiltonian; the second term in (1), written as

$$\hat{H}_{dd} = \sum_{i<k} D_{ik} [\vec{S}_i \vec{S}_k - 3 r_{ik}^{-2} (\vec{r}_{ik} \vec{S}_i)(\vec{r}_{ik} \vec{S}_k)] \qquad (3)$$

corresponds to the dipole-dipole interaction; and the third term

$$\hat{H}_{anis} = \sum K_{ik} r_{ik}^{-4} (\vec{S}_i \vec{r}_{ik})^2 (\vec{S}_k \vec{r}_{ik})^2 \qquad (4)$$

takes into account the magnetic anisotropy. In (2)-(4), $\vec{S}_i$, is the vector operator of the spin of the $i$th atom measured in $\hbar$ units; $\vec{r}_{ik}$ is the position vector connecting the $i$th and $k$th atoms; $A_{ik}$, the exchange integrals; $D_{ik}$, the parameters of the dipole-dipole interaction; and $K_{ik}$, the parameters of quadrupole-quadrupole interaction. As was indicated by Van Vleck [11], the simplest microscopic model that allows for magneto-crystalline anisotropy suggests that the spin-spin interaction is of the quadrupole character.

Let us analyze the symmetry properties of the magnetic Hamiltonian. It is obvious that the first term is invariant with respect to the simultaneous rotation of all spins through the same angle. Moreover, under the action of operations of the symmetry group of a paramagnetic crystal, the atoms will move into crystallo-graphically equivalent positions. The exchange integrals $A_{ik}$ only depend on the absolute value of the position vector $\vec{r}_{ik}$ that connects the interacting atoms and do not change upon transformations caused by the symmetry elements of the space group $G$ of the paramagnetic crystal lattice. Thus, the symmetry group of the Hamiltonian (2) is the exchange paramagnetic group $G \times O(3)$, where O(3) is the three-dimensional group of spin rotations. The $G \times O(3)$ group is a color symmetry group of the so-called $P$ type [12]. The second and third terms are related to relativistic interactions and have a lower symmetry as compared to that of the exchange Hamiltonian. This symmetry coincides with the symmetry of the crystal lattice of the paramagnetic crystal, which is described by the space group $G$. As can be seen from (3) and (4), the rotating part of the symmetry element must act not only on the position vector $\vec{r}_{ik}$ but also on the spin vectors $\vec{S}_i$, and $\vec{S}_k$. As to pure translations, both (3) and (4) are obviously invariant with respect to them, since the parameters of the dipole-dipole *($D_{ik}$)* and quadrupole-quadrupole *($K_{lk}$)* interaction depend only on the spacing $\vec{r}_{ik}$ between the spins. Moreover, expressions (3) and (4) are invariant with respect to the operation of inversion of the spin direction $\vec{S}_i \to -\vec{S}_i$. This means that the symmetry group of the total Hamiltonian is the Shubnikov paramagnetic group $G1' = G \times \{1,-1\}$.

The energy spectrum of a macroscopic body is quasi-continuous [10]; therefore, at the Curie point $T_c$, states that are associated with zero magnetization vectors neighbor with states that are associated with nonzero magnetization. The symmetry of the magnetic Hamiltonian and the symmetry of the statistical-equilibrium state above the Curie point are the same, whereas below the Curie temperature, the symmetry of the statistical-equilibrium state becomes lower than the symmetry of the magnetic Hamiltonian. Such thermodynamic systems are called systems with a spontaneous symmetry breakdown [13]. Above the Curie point, the convenient Bogolyubov averages and quasi-averages coincide, whereas below the Curie point, this is not the case [13].

It is commonly accepted that the symmetry of the total magnetic Hamiltonian (1) breaks down spontaneously at the Curie point. However, the hypothesis on the existence of spin-orbit phase transitions [4-6] states that the spin-orbit interaction occurs at a temperature $T_{ls} < T_c$. This hypothesis is, thus, equivalent to the following statement: at the temperature $T_c$, there occurs a spontaneous breakdown of the exchange Hamiltonian (2), whereas at the temperature $T_{ls} < T_c$, there occurs a spontaneous breakdown of the total Hamiltonian, i.e., a spin-orbit (relativistic) phase transition takes place. Below the Curie point, there arises a spontaneous magnetization vector $\vec{M}$, whose orientation relative to the crystallographic axes is arbitrary, i.e., the resultant ferromagnetic phase is isotropic. Below $T_{ls}$, the relativistic phase transition leads to the appearance of a magnetic anisotropy and of magnetostrictive distortions of the crystal, i.e., the resultant ferromagnetic phase is anisotropic.

A question that arises is in which crystals we should expect the existence of an isotropic phase. The main carriers of magnetic properties in crystals are the elements of the iron (3$d$) and rare-earth (4$f$) groups. In the former, the $LS$ coupling arises, whereas in the latter, this is the $jj$ coupling. The electron cloud of a *3d* ion is spherically symmetric, and the cloud shape has no significant correlation with the spin direction. In the 4$f$ ion, on the contrary, the electron cloud is strongly anisotropic ("rigid"), and there exists a strong correlation between the distribution of the electric charge and the direction of the magnetic moment of the ion. For this reason, the magnetic symmetries of the exchange and nonexchange magnetic materials should be described differently even in the paramagnetic phase. The exchange magnetic materials that contain magnetically active *3d* ions are invariant in the paramagnetic phase with respect to the group of three-dimensional spin rotations O(3), because for a spin that is independent of the distribution of the electric charge, no potential barriers exist that separate one spin direction from any other. If the electron cloud is anisotropic, then there are several states (orientations of the electron cloud) in the center of a coordination polyhedron that correspond to the minimum energy of the electrostatic interaction. A fixed position of the anisotropic electron cloud corresponds to a fixed orientation of the total magnetic moment of the ion $\vec{j}$. In fact, no such static picture is realized; rather, there takes place a dynamic effect, i.e., continuous transitions occur between all available states of the ion; note that the number of these states is finite. Exchange magnets also exhibit transitions between possible states, but these states form a two-dimensional continuum equivalent to the surface of the unit sphere.

The concept of spin-orbit phase transitions was introduced in [4-6] and some other works to explain some experimental situations; we believe, however, that these situations are unconvincing; moreover, their discussion was purely qualitative. Therefore, the question of whether or not spin-orbit phase transitions exist in reality and how they can be unambiguously identified remains open. As a possible experimental support for the existence of spin-orbit phase transitions and of an isotropic magnetic phase, we consider the behavior of various physical characteristics of copper ferrite ($CuFe_2O_4$) crystals.

### 3. Experimental data and selection of a model

The copper ferrite $CuFe_2O_4$ has been studied starting from the beginning of the 1960s (see [14-20] and references to chapter 29 of [14]). The main interest in this compound is the structural phase transition existing in it, which is usually associated with the cooperative Jahn-Teller effect. The transition occurs as the temperature decreases to $T_{JT} \approx 631\,\text{K}$ [5] and is accompanied by the lowering of the cubic symmetry of its crystal lattice to tetragonal and an anomalous behavior of the magnetic and some other physical properties [15-20]. An interesting feature of the Jahn-Teller effect in copper ferrite is the fact that it is observed in the magnetically ordered phase: at $T_c = 730$ K [14], there occurs a ferromagnetic phase transition in the crystal.

Copper ferrite has the spinel structure, whose symmetry at sufficiently high temperatures is characterized by the space group $O_h^7$. An analysis of the experimental data given in [14-20] suggests that in the temperature range of 630-730 K (1) the crystal lattice has a cubic symmetry characteristic of the paramagnetic phase; (2) the magnetic anisotropy of the crystal is anomalously small; and (3) the ferromagnetic structure is collinear, of the Neel type. With the transition to the tetragonal phase (at $T < T_{JT}$), the magnetic anisotropy increases sharply and the magnetization suffers a small jump in the vicinity of the transition point. All these properties suggest, that copper ferrite is a typical exchange magnet, and the "tetragonal" phase transition is the transition from an isotropic to an anisotropic ferromagnetic phase induced by the cooperative effect of relativistic interactions.

Let us now turn to the construction of a theoretical model which would consistently describe the behavior of the copper ferrite crystal in a sufficiently wide temperature range from slightly above the Curie point to room temperature and below. We will neither discuss nor criticize theoretical works devoted to the Jahn-Teller effect, because we set a quite different task of including the structural transition associated with this effect into the general picture of phase transitions occurring in the ferrite with the spinel structure ($CuFe_2O_4$) rather than of explaining and describing the Jahn-Teller effect. In our opinion, such a picture is given by the thermodynamic Landau theory [10], which has been developed in much detail, sufficiently to allow for the magnetic and crystal-chemical symmetry of the lattice; this theory proceeds from the fundamental principles of statistical thermodynamics and permits one to sufficiently easily describe experimental results.

To describe the whole body of experimental data according to the Landau theory, one should, first, select the symmetry group of the initial (high-symmetry) phase and, second, choose a critical order parameter. By the critical representation, it is usually meant an irreducible representation of the symmetry group of the initial phase which determines the lowering of symmetry at the phase transition point and according to which there occurs transformation of the critical order parameter. After this, all subsequent actions are performed by the standard scheme which is described, e.g., in [21,22]. As we have already seen in Section 2, we should select the exchange paramagnetic group as the symmetry group of the paramagnetic phase for exchange magnets. For the ferrite with spinel structure, this will be the $O_h^7 \times O(3)$ group. The transition to the ferromagnetic phase is accompanied by the development of a spontaneous magnetization and, consequently, the role of the critical order parameter will be played by the magnetization vector $\vec{S}$. We designated the magnetization vector by the letter $\vec{S}$ in order to emphasize its purely spin origin in the isotropic phase. If this vector is considered as a stationary matrix (see below) in the spirit of [23], it is associated with a stationary matrix with one column and three rows

$$\vec{S} = \begin{pmatrix} S_1 \\ S_2 \\ S_3 \end{pmatrix}$$

The three components form a vector with respect to the group of three-dimensional spin rotations O(3), and each of these components is a scalar with respect to the transformations due to the elements of the $O_h^7$ group. Therefore, the critical representation is the $A_{1g} \times V'$ irreducible representation (IR) of the magnetic group $O_h^7 \times O(3)$, where $A_{1g}$ is the unit IR of the cubic group $O_h^7$, and $V'$ is the pseudo-vector IR of the O(3) group. The thermodynamics of the transition according to the $A_{1g} \times V'$ IR was described in detail in [24], so that we will not consider this question and proceed with the description of the transition from the isotropic ferromagnetic phase into possible anisotropic phases.

### 3. Determining symmetry groups of anisotropic ferromagnetic phases

Upon the transition from the paramagnetic phase into the isotropic ferromagnetic phase, the symmetry is lowered as follows: $O_h^7 \times O(3) \to O_h^7 \times SO(1)$ (at $T = T_c$), i.e., the cubic symmetry of the crystal lattice is retained and the magnetic symmetry of the crystal is lowered. The three-parameter group of improper spin rotations transforms into a single-parameter group of proper rotations about the direction of the magnetization vector $\vec{S}$, and the operation of inversion of the spin direction is impossible, since the phase is magnetically ordered. In order to describe the transition from the isotropic into the anisotropic phase, we cannot construct a new theoretical scheme and conjecture what will be the new critical irreducible representation. Following our main premise, we assume that the critical irreducible representation of the $O_h^7$ group, which induces the transitions from the isotropic into possible anisotropic phases, should unambiguously be deduced from the critical $A_{1g} \times V'$ IR of the $O_h^7 \times O(3)$ group. Such a group-theoretical operation exists and is called the restriction of the IR of the direct product of two groups to one of the multipliers [12]. Let us designate the restriction of the $A_{1g} \times V'$ IR to the $O_h^7$ group by the same symbol but taken in brackets, i.e., $\lfloor A_{1g} \times V' \rfloor$. In the general case, such a restriction is reducible and is the direct sum of other IRs of the same space group belonging to the star of the same wave vector as the first multiplier and having the same parity relative to the spatial inversion [25]. This restriction is called the exchange multiplet [12]. Calculations using the general group-theoretical formulas lead to the following result: $\lfloor A_{1g} \times V' \rfloor = F_{1g}$; that is, the exchange multiplet degenerates to a singlet, and we have no problem of selecting the critical IR. Thus, the critical IR that induces phase transitions from the isotropic ferromagnetic phase to anisotropic ferromagnetic phases is the three-dimensional pseudo-vector irreducible representation $F_{1g}$ of the cubic space group $O_h^7$.

**Table 1.** Low-symmetry phases induced by the IR $F_{1g}$, of the $O_h$ group.

| $\vec{c}$ | $ccc$ | $occ$ | $coo$ | $c_1 c_2 c_3$ |
|---|---|---|---|---|
| $G_D$ | $C_{3i}^2$ | $C_{2h}^3$ | $C_{4h}^6$ | $C_i^1$ |

Let us now determine the magnetic groups of symmetry of the anisotropic phases. This may be done using a purely algebraic method [21,22], which consists in the following. For the three-dimensional IR $F_{1g}$, a set of matrices of the representation $M(g)$ is specified, where $g$ are the elements of the $O_h^7$ group and $\vec{c}$ is assumed to be an order parameter that is transformed through the IR $F_{1g}$. We may believe that the order parameter $\vec{c}$ is a vector of some space of a proper dimension. This space is called the image space [26]. The arrangements of the order parameter in the image space that differ in symmetry are associated with different subgroups of the $O_h^7$ group. These can be found by solving the algebraic equations $M(g)\vec{c} = \vec{c}$ for the matrices of the representation. The vector that remains unaltered upon the action of the matrix on it is called the stationary vector of this matrix. For each subgroup of the initial group, there exists its own stationary vector. In order to find this vector, it is not necessary to use all the matrices of the representation, but it is sufficient to use the generating matrix. The calculations for the $O_h^7$ group were performed in [21] and the results that are necessary for our work are given in Table 1.

Now, we may turn to the calculation of the temperature dependences of the physical characteristics of ferrites.

## 4. Temperature dependence of magnetic anisotropy constants

In order to obtain the temperature dependences of magnetic anisotropy constants, the thermodynamic Landau potential of a cubic crystals should be constructed. It is well known that the Landau potential should depend on invariants that consist of the components of the order parameter $\vec{c}$. The IR $F_{1g}$ in the space of the order parameter is associated with the point group $O$, which is called the image group [26]. All invariant polynomials composed of the components of the order parameter are expressed through several polynomials that form an integer rational basis of invariants (IRBI). For the $L$ group coinciding with the point group $O$, the IRBI consists of the following three polynomials [26]:

$$I_1 = c_1^2 + c_2^2 + c_3^2; I_2 = c_1^2 c_2^2 + c_2^2 c_3^2 + c_3^2 c_1^2; I_3 = c_1^2 c_2^2 c_3^2 \qquad (5)$$

Apart from these three polynomials, the IRBI also contains a polynomial of the ninth order. If we can restrict ourselves to the Landau potential that only takes into account changes in two control parameters, e.g., the temperature and the external magnetic field, when considering the thermodynamics of the transition to the anisotropic magnetic phase, then we may restrict ourselves to a potential of the sixth degree (see below), and we will not need the ninth-degree polynomial. It is desirable that the Landau potential be reduced to the form that is usually employed in an analysis of spin-reorientational transitions. To this end, we express the order parameter $\vec{c}$ through the unit vector $\vec{m}$ by the formula $\vec{c} = c\vec{m}$. In this case, the basis invariants acquire the following form:

$$I_1 = c^2; I_2 = c^4 s; I_3 = c^6 m_1^2 m_2^2 m_3^2 , \tag{6}$$

where $s = m_1^2 m_2^2 + m_2^2 m_3^2 + m_3^2 m_1^2$. As to the Landau potential, we only know that it is an invariant function of the order parameter. In addition, it is known that it has a critical point, which is degenerate in the general case. An arbitrary smooth k-parameter function may be represented in the vicinity of a degenerate critical point in the so-called normal form, i.e., in the form of a polynomial of a finite degree, and the procedure of reducing to the normal form at a given number of control parameters is quite unambiguous [27]. If there are only two governing parameters, e.g., temperature $T$ and pressure P, then the normal form of the Landau potential is a polynomial of the sixth degree

$$F(\vec{c}) = a_1 I_1 + a_2 I_1^2 + a_3 I_1^3 + b_1 I_2 + b_2 I_3 + b_3 I_1 I_2 . \tag{7}$$

Substituting the explicit expressions for the invariants into (7), we obtain

$$F(\vec{c}) = a_1 c^2 + a_2 c^4 + a_3 c^6 + (b_1 c^4 + b_3 c^6)s + b_2 c^6 m_1^2 m_2^2 m_3^2 \tag{8}$$

Let us compare our potential with the classical potential that takes into account only the anisotropy energy of a cubic crystal

$$U_{anis} = K_1(T)s + K_2(T) m_1^2 m_2^2 m_3^2 . \tag{9}$$

where $K_1(T)$ and $K_2(T)$ are the first and second constants of the magnetocrystalline anisotropy of cubic crystals. Comparing (9) with the two last terms of (8), we find expressions for these anisotropy constants through the absolute value of the order parameter

$$K_1(c) = b_1 c^4 + b_3 c^6; K_2(c) = b_2 c^6 , \tag{10}$$

Given the temperature dependence of the order parameter c = c(T), formulas (10) yield the sought-for temperature dependence of the anisotropy constants. Since $c \approx \sqrt{T_{ls} - T}$ in the Landau theory, we have

$$K_1 \approx (T_{ls} - T)^2; K_2 \approx (T_{ls} - T)^3 \tag{11}$$

Note that the above formulas for the temperature dependences are valid near the isotropic-anisotropic phase transition temperature $T_{ls}$. Formulas (11) remain valid also for nonexchange ferromagnets, in which the isotropic phase does not exist. In this case, the $T_{ls}$ temperature should be replaced by the Curie temperature $T_c$. Formulas (10)-(11) give zero values of magnetic anisotropy constants at $T > T_{ls}$ in agreement with experiment. Note that our formulas differ from the classical formulas of Akulov-Zener, which express the magnetic anisotropy constants through the magnetization

$$K_1 \cong M^{10}; K_2 \cong M^{21} . \tag{12}$$

When applied to nonexchange ferromagnets, formulas (11) yield

$$K_1 \cong M^4; K_2 \cong M^6$$

Let us make a remark concerning the physical meaning of the order parameter $c$ that describes the phase transition from the isotropic magnetic phase into the anisotropic phase. Since the magnetizations of the sublattices in the isotropic phase differ from zero and have finite values, whereas the order parameter should be zero in the isotropic phase, the order parameter is not equal to the magnetization vector $\vec{S}$. In fact, this is the orbital contribution to the vector of spontaneous magnetization caused by the cooperative effect of relativis-tic interactions, and the total magnetization is equal to the sum of terms of the exchange and relativistic origin $\vec{M} = \vec{S} + \vec{c}$. The appearance of this contribution causes both the changes in the magnitudes of the magnetizations of the sublattices and the rotation of the magnetization vector from the arbitrary direction along one of the easy axes. The absolute value of the order parameter $c$ gives the value of that small jump of the magnetization that is observed experimentally near the tetragonal phase transition.

## 5. Physical effects that accompany the isotropic-anisotropic phase transition

Upon the phase transition to an anisotropic phase, not only the magnetic state changes, but various accompanying effects take place as well. Let us consider them from the general group-theoretical viewpoint, following [21,22]. As we have already said above, the $F_{1g}$ irreducible representation is "principal" in some way, since it is this IR that determines the decrease in crystal symmetry at the phase transition. In the immediate vicinity of the transition point, effects that are due to the critical IR are most pronounced (small to the first order in c). However, the phase transition is accompanied by a number of physical changes that are compatible with the symmetries of the newly forming phases. To each of these changes, there corresponds its own IR, which is called noncritical or associated [21]. The critical and noncritical IRs are determined in terms of linear algebra quite unambiguously and, taken jointly, form the so-called condensate. The information concerning the condensate of the IR $F_{1g}$ which we need is contained in [28], and the results are given in Table 2. Note that the IR $F_{1g}$ itself also enters into the condensate as a secondary IR, although this is not shown in Table 2. The components of the critical and secondary parameters may be used to form mixed invariants. In the mixed invariants, the components of the secondary order parameters are contained to the first degree, and the components of the critical parameter to higher degrees. These degrees are given in Table 3. The symbol "+" in the table indicates the presence, and the symbol "-" means the absence of the corresponding invariant in the Landau potential.

**Table 2.** Total condensate of stationary vectors of the critical IR $F_{1g}$,

belonging to the star of the wave vector $\vec{k} = 0$ of space group $O_h^7$

| $\vec{c}$ | $G_D$ | $A_{1g}$ | $A_{2g}$ | $E_g$ | $F_{2g}$ |
|---|---|---|---|---|---|
| $ccc$ | $C_{3i}^2$ | $a$ | $a$ | – | $aaa$ |
| $occ$ | $C_{2h}^3$ | $a$ | – | $ao$ | $aab$ |
| $coo$ | $C_{4h}^6$ | $a$ | – | $a, -\sqrt{3}a$ | – |
| $c_1 c_2 c_3$ | $C_i^1$ | $a$ | $a$ | $a, b$ | $abc$ |

**Table 2.** Entering of the noncritical IRs of space group $O_h^7$ into the direct symmetric degrees of the critical IR $F_{1g}$.

| $m_s$ | $A_{1g}$ | $A_{2g}$ | $E_g$ | $F_{1g}$ | $F_{2g}$ |
|---|---|---|---|---|---|
| 2 | + | – | + | – | + |
| 3 | + | + | + | + | + |
| 4 | + | – | + | + | + |
| 5 | + | + | + | + | + |

Let us now clarify the physical meaning of the concrete degrees of freedom associated with each secondary order parameter. To the unit IR $A_{1g}$, there corresponds isotropic relativistic magnetostriction. The corresponding secondary order parameter is equal to the sum of the diagonal elements of the deformation tensor $a = u_{xx} + u_{yy} + u_{zz}$. The components of the deformation tensor $u_{ij}$ may form two more secondary order parameters, namely, a two-dimensional order parameter with components

$$a_1 = \frac{1}{\sqrt{6}}(2u_{zz} - u_{xx} - u_{yy}); a_2 = \frac{1}{\sqrt{2}}(u_{yy} - u_{zz}) \quad (13)$$

which is transformed through the IR $E_g$, and an order parameter

$$a_1 = u_{yz}; a_2 = u_{zx}; a_3 = u_{xy} \quad . \quad (14)$$

which is transformed via the three-dimensional IR $F_{2g}$.

The two-dimensional order parameter (13) describes compressive or tensile deformations, and the three-dimensional order parameter (14) describes shear deformations. For describing the relativistic magnetostriction, the Landau potential (8) should be augmented with mixed invariants composed of the components of the secondary order parameters (13)—(14) and with quadratic invariants of the same parameters. To facilitate comparison with the commonly used designations [29], we write the elastic and the magnetoelastic energies explicitly through the components of the deformation tensor

$$F_y = \frac{C_{11}}{2}(u_{xx}^2 + u_{yy}^2 + u_{zz}^2) + \frac{C_{44}}{2}(u_{xy}^2 + u_{yz}^2 + u_{zx}^2) + C_{12}(u_{xx}u_{yy} + u_{yy}u_{zz} + u_{zz}u_{xx}), \quad (15)$$

$$F_{my} = \alpha_1 c^2 [m_x^2 u_{xx} + m_y^2 u_{yy} + m_z^2 u_{zz} - \frac{1}{3}(u_{xx} + u_{yy} + u_{zz})] + \\ + 2\alpha_2 c^2 (m_x m_y u_{xy} + m_y m_z u_{yz} + m_z m_x u_{zx}) + \alpha_3 c^4 (u_{xx} + u_{yy} + u_{zz})s \quad (16)$$

In the above formulas, $C_{ij}$ are the elastic constants of the cubic crystal; $\alpha_i$, are the magnetoelastic constants; and $m_x, m_y$, and $m_z$ are the components of the unit vector directed along the antiferromagnetism vector. Note that, unlike the commonly accepted approach [29], we separated the temperature dependences from the magnetoelastic constants using the factors $c^2$ and $c^4$. By differentiating the sum of the elastic and magnetoelastic energy with respect to the components of the deformation tensor and equating the derivative to zero, we obtain the dependence of the components of the tensor of relativistic magnetostriction on the components of the order parameter

$$u_{ik} \cong c^2 m_i m_k \cong (T_{ls} - T) m_i m_k \quad (17)$$

For nonexchange ferromagnets, the temperature of the relativistic transition $T_{ls}$ in (17) should be replaced by the Curie temperature $T_c$.

**Table 4.** Vector basis for the positions 32(*e*) of the space group $O_h^7$.

| IR | 1 | 2 | 3 | 4 | 5 | 6 | 7 | 8 |
|---|---|---|---|---|---|---|---|---|
| $E_g$ | $11\bar{2}$ $\bar{e}eo$ | $1\bar{1}2$ $\bar{e}\bar{e}o$ | $\bar{1}\bar{1}2$ $e\bar{e}o$ | $\bar{1}12$ $eeo$ | $\bar{1}12$ $eeo$ | $\bar{1}\bar{1}2$ $e\bar{e}o$ | $1\bar{1}2$ $\bar{e}\bar{e}o$ | $112$ $eeo$ |
| $F_{1g}$ | $1\bar{1}0$ $\bar{1}01$ $01\bar{1}$ | $\bar{1}\bar{1}0$ $101$ $0\bar{1}1$ | $\bar{1}10$ $10\bar{1}$ $0\bar{1}1$ | $110$ $\bar{1}0\bar{1}$ $0\bar{1}1$ | $110$ $\bar{1}0\bar{1}$ $0\bar{1}\bar{1}$ | $\bar{1}10$ $\bar{1}0\bar{1}$ $011$ | $\bar{1}\bar{1}0$ $\bar{1}01$ $011$ | $1\bar{1}0$ $101$ $0\bar{1}\bar{1}$ |

## 6. Analysis of the structural changes in copper ferrite

Below the temperature of the Iahn-Teller (in fact, the relativistic) transition in the copper ferrite, tetragonal distortions of the crystal lattice are observed, and the tetragonal axis becomes the easy axis of magnetization. X-ray diffraction and other structural studies (whose results are given in [15-20]) suggest that the tetragonal phase has a symmetry described by the space group $D_{4h}^{19}$. However, it follows from our theory that the tetragonal phase should have a symmetry that is characterized by the space group $C_{4h}^{6}$. Thus, the theory appears to contradict the experiment. We show that this contradiction is only apparent.

The space group $D_{4h}^{19}$ appears if the phase transition occurs through the irreducible representation $E_g$ of the space group $O_h^7$. We believe, however, that the critical irreducible representation is the pseudovector irreducible representation $F_{1g}$ resulting in a tetragonal phase with symmetry $C_{4h}^6$. The nature of displacements for the IRs $E_g$ and $F_{1g}$ is quite different, which can be seen from the calculations performed by the formula

$$\vec{u}_i(\vec{r}) = \sum_{\alpha} c_\alpha \vec{\varphi}_{\alpha i}(\vec{r}) \, , \tag{18}$$

where $\vec{u}_i(\vec{r})$ the displacement of the ith atom in the unit cell of the spinel structure; and $\vec{\varphi}_{\alpha i}$ are the basis functions of the representation considered. For both IRs, only the 32(e) positions occupied by oxygens suffer displacements. This can be expressed in the group-theoretical language as follows: the IRs $E_g$ and $F_{1g}$, enter into the mechanical representation at positions 32(e). In the cubic unit cell (Fig. la), there are eight oxygen positions, which are rigorously enumerated (Fig. lb). The vector basis functions, which were borrowed from [30], are given in Table 4. The symbol "e" in the table denotes the number $\sqrt{3}$, and the bar over the symbol means that the number is negative. The transition into the tetragonal phase through the IR $E_g$ is associated with the stationary vector (C, 0), and the transition through the IR $F_{1g}$ with the stationary vector (C, 0, 0). If we substitute the basis functions from Table 4 and the stationary vectors (C, 0) and (C, 0, 0) into (18), we will see that the displacements of each of the eight oxygen atoms are only determined by the first line of the vector basis for both the $E_g$ and $F_{1g}$ IRs.

Figure la shows a portion of the unit cell of the spinel structure which is equivalent to its primitive unit cell. The A atoms, which occupy the tetrahedral positions 8(*a*) are shown as stroked circles. The B atoms, which occupy the octahedral positions 16(*d*), are shown by the solid circles, and the oxygen atoms are given by the empty circles. Figure lb displays the upper projection of this unit cell (only oxygen atoms are shown). Figure schematically shows the anion displacements for the $E_g$ and $F_{1g}$ IRs. These figures correspond to Fig. 1b. which also shows the oxygen anion numbers.

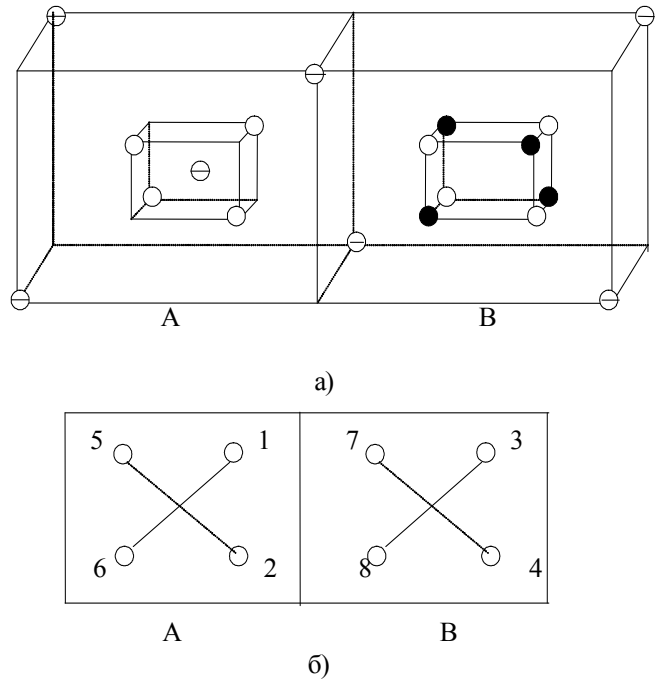

**Fig. 1.** A portion of the unit cell of the spinel structure.

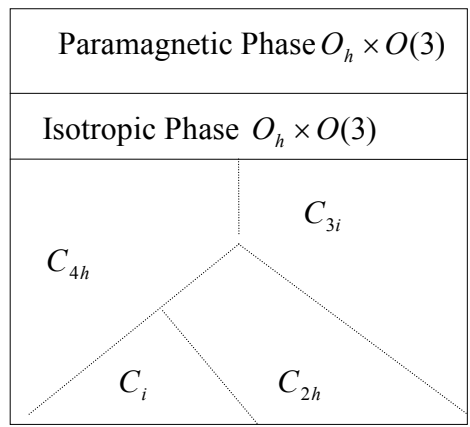

**Fig. 2.** The diagram of possible magneto-crystalline states of a cubic ferromagnet

Figure 2 displays the diagram of possible magneto-crystalline states of a cubic ferromagnet, which contains paramagnetic, isotropic, and anisotropic phases (we gave the symbols of point groups instead of the total symbols of space groups). Solid lines correspond to first-order transitions. The single-parameter phases $C_{4h}$ and $C_{3i}$ always neighbor with the paramagnetic phase, so that the transition of the crystal into the ferromagnetic state is accompanied by the appearance of an easy axis [100] or [111]. The transition into these states is a second-order phase transition. Under certain conditions, the paramagnetic phase may also neighbor with a single-parameter phase $C_{2h}$, which has an easy axis of the [011] type.

In this case, the boundary between the $C_{4h}$ and $C_{3i}$ phases degenerates into a tetra-critical point, at which the paramagnetic and three single-parameter phases meet. The dashed lines between the regions of existence of the ferromagnetic phases correspond to the lines of spin-reorientation transitions, which are first-order phase transitions. For example, when the representative point crosses the boundary between the tetragonal phase $C_{4h}$ and the triclinic phase $C_i$ a spin-reorientation transition into a canted phase with an easy axis of the [lmn] type occurs.

If we take a solid solution of two magnetic compounds that have the same crystal-lattice type but different easy-axis types below the relativistic phase transition temperature $T_{ls}$, then a change in the concentration of one of the components may result in a spin- reorientation transition of the [111] ↔ [100] type. Such a transition can occur at temperatures sufficiently close to the $T_{ls}$ point. At lower temperatures, this transition occurs through an intermediate phase: [111] ↔ [011] ↔ [100].

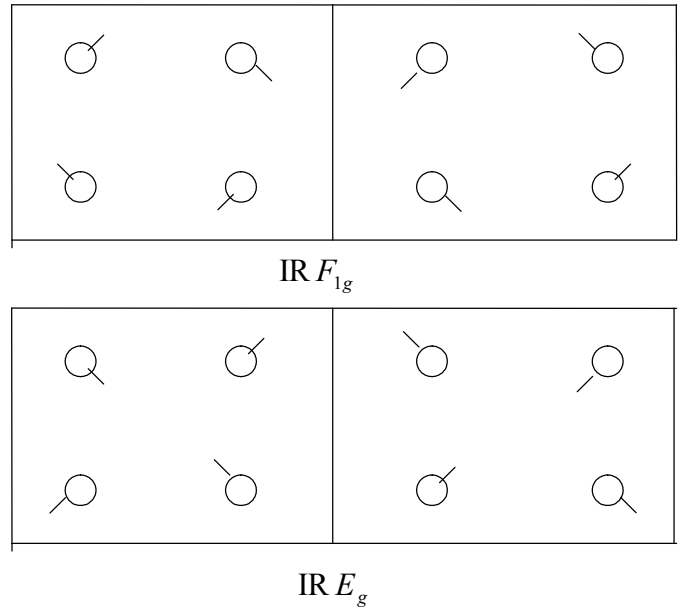

IR $F_{1g}$

IR $E_g$

**Fig. 3.** The character of displacements for two IRs

It can be seen from Fig. 3 that the character of displacements for these two IRs is quite different. How can we explain that the pattern of displacements that are observed experimentally corresponds to the IR $E_g$, whereas the critical irreducible representation is $F_{1g}$. The IR $E_g$ enters into the total condensate of the critical IR $F_{1g}$ as a secondary IR. The relative contributions of the critical and noncritical displacements to the final pattern of displacements in the vicinity of a phase transition point are different; namely, the critical displacements are large compared to the noncritical ones. Far from the phase transition point, the noncritical displacements can exceed the critical displacements and the X-ray diffraction will show the pattern of displacements corresponding to the space group $D_{4h}^{19}$, although the true symmetry group of the crystal lattice is $C_{4h}^{6}$ (see [21] for more detail).

## 6. Conclusion

From the above, the following conclusion may be drawn: we indeed succeeded in theoretically and experimentally substantiating the existence of spin-orbit (relativistic) phase transitions. The meaning of the above-said consists in the following. We developed a phenomenological theory of relativistic phase transitions in terms of the thermodynamic Landau theory. The main postulate of the Landau theory, i.e., the statement that the symmetry groups of low-symmetry phases are subgroups of the symmetry group of the high-symmetry phase, remained hard-and-fast. But the idea that the low-symmetry phases correspond to the absolute minima in the nonequilibrium potential which is invariant with respect to the symmetry group of the high-symmetry phase had to be sacrificed. In this case, the exchange paramagnetic group $G \times O(3)$ should be selected as the symmetry group of the high-symmetry (paramagnetic) phase. The absolute minima of the non-equilibrium thermodynamic potential that is invariant relative to this group correspond to different isotropic magnetically ordered phases. In order to consider transitions from the isotropic into anisotropic phases, we should construct a new potential which is invariant relative to only the space group. The critical irreducible representation of group $G$ that induces the transitions into different anisotropic phases enters into the exchange multiplet generated by the critical irreducible representation of the $G \times O(3)$ group that induces phase transitions at the Curie point. This ideology was illustrated using the copper ferrite $CuFe_2O_4$, which is ferromagnetic. Our theory permits us to describe changes in the magnetic and crystallographic symmetry at the Curie point and at the point of the tetragonal phase transition, as well as changes of some physical properties.

Note that the attempt to retain the Jahn-Teller interpretation of the tetragonal phase transition at $T_{JT} \approx 631~K$ leads to serious difficulties. Indeed, as was said in [24, p. 203], "when magnetized along the direction of the cubic unit-cell edge, a cubic crystal becomes weakly tetragonal, while magnetized along the spatial diagonal of the unit cell cube, it becomes rhombohedral." This means that if we assume that the easy axes appear directly at the Curie point, the crystal of the copper ferrite indeed becomes tetragonal already at the Curie point $T_c$ = 730 K, while at $T_{JT} \approx 631 K$ there occurs an isostructural phase transition with a sharp increase in the degree of tetragonality due to the cooperative Jahn-Teller effect. But, with this approach, it is difficult to find a simple and clear link between the changes in the symmetry and the magnetic and other physical properties. The attempt to discard the general scheme of the Landau theory and to introduce the "foreign" idea of the Jahn-Teller effect into the picture of magnetostructural changes in the crystal with the only purpose of not recognizing the existence of the isotropic magnetic phase is rather eclectic, in our opinion. The situation resembles the events of the last quarter of the last century, when various arguments were given in order to retain the mechanical supports for the Maxwell field equations and not to recognize the independent existence of the field.

Thus, the $T_{JT}$ temperature ($\approx 631K$) is most likely to be the temperature of the relativistic phase transition ($T_{ls}$) in the copper ferrite, while in the temperature range of 630 to 730 K, the magnetocrystalline constants appear to be exactly zero.

———————————————